\documentclass[11pt]{article}

\usepackage[sort,compress]{cite}

\usepackage{color}

\definecolor{grey}{rgb}{0.4,0.4,0.5}
\definecolor{darkgreen}{rgb}{0,0.5,0}
\definecolor{darkred}{rgb}{0.6,0.0,0}
\definecolor{lightbrown}{rgb}{1,0.9,0.8}
\definecolor{brown}{rgb}{0.6,0.3,0.3}
\definecolor{darkblue}{rgb}{0,0,0.8}
\definecolor{darkmagenta}{rgb}{0.5,0,0.5}
\definecolor{myurlcolor}{rgb}{0.4,0,0}
\definecolor{mycitecolor}{rgb}{0,0.4,0}
\definecolor{mylinkcolor}{rgb}{0,0,0.4}

\usepackage[colorlinks=true, urlcolor = myurlcolor, citecolor = mycitecolor, linkcolor= mylinkcolor, bookmarks=false]{hyperref}
\usepackage{amsfonts,amsmath,amssymb}
\usepackage{graphicx}
\usepackage{dsfont}

\usepackage{arydshln}

\usepackage{xspace}

\usepackage{relsize}
\usepackage{color}
\usepackage{braket}

\usepackage{bm}

\usepackage[bottom]{footmisc}

\usepackage{stmaryrd}
\usepackage{setspace}

\usepackage{nicefrac}

\newcommand{\email}[1]{\href{mailto:#1}{#1}}

\newcommand{\namedref}[2]{\hyperref[#2]{#1~\ref*{#2}}}

\def\half{\nicefrac{1}{2}}
\def\ii{\mathrm i}
\def\dd{\mathrm d}
\def\pa{\partial}

\def\da{\dagger}

\def\vol{\mathcal V}

\def\la{\label}

\def\be{\begin{equation}}
\def\ee{\end{equation}}

\newcommand{\bei}{\begin{itemize}}
\newcommand{\eei}{\end{itemize}}

\newcommand{\bee}{\begin{enumerate}}
\newcommand{\eee}{\end{enumerate}}

\newcommand{\Der}[2]{\frac{\dd #1}{\dd #2}}
\newcommand{\pDer}[2]{\frac{\pa #1}{\pa #2}}

\def\a{\alpha}

\def\k{\kappa}
\def\lam{\lambda}

\def\s{\sigma}

\def\th{\theta}

\def\OH{\bm H}
\def\OA{\bm A}

\def\OV{\bm V}

\def\Os{{\bm s}}

\def\OT{\bm T}

\def\Oet{{\bm \eta}}

\def\Oc{\bm c}
\def\Ocd{\bm c^\da}

\def\Oq{\bm q}
\def\Oqd{\bm q^\da}

\def\On{{\bm n}}

\def\bOn{\bar{\bm n}}

\def\Oth{\bm \th}

\def\bs{{\bar{\sigma}}}

\def\bsu{\uparrow}
\def\bsd{\downarrow}

\def\ssu{{\mathsmaller{\uparrow}}}
\def\ssd{{\mathsmaller{\downarrow}}}

\def\qe{{\circ}}
\def\qd{{\bullet}}
\def\bqe{{\mathlarger{\mathlarger{\circ}}}}
\def\bqd{{\mathlarger{\mathlarger{\bullet}}}}

\def\da{\dagger}

\def\st{\star}

\def\hst{\hat \star}
\def\cst{\check \star}

\def\bs{\bar{\sigma}}

\def\ssu{\uparrow}
\def\ssd{\downarrow}

\def\qe{{\circ}}
\def\qd{{\bullet}}

\def\B{\ddag}

\begin{document}

\mbox{}


\medskip

\begin{center}{\large \textbf{
Hidden structure in the spectra of strongly correlated electrons in 1d
}}\end{center}

\begin{center}
\textsc{Eoin Quinn}
\end{center}

\begin{center}
{\it Institute for Theoretical Physics, University of Amsterdam, Science Park 904,\\ 
1090 GL Amsterdam, The Netherlands}
\\ 
\bigskip
{\small\ttfamily\email{epquinn@gmail.com}}\\
\bigskip
\today
\end{center}

\bigskip

\begin{center}
\textbf{Abstract}\\
\vspace{5mm}
\begin{minipage}{12.5cm}
We identify a structure in the spectra of 1d lattice models of interacting electrons,
characterised by an anomalous gapped branch of elementary excitations. Focusing on a family of Bethe ansatz solvable models, where all excitations are stable against decay, we make a four-way classification of the energy spectrum along with a model belonging to each class. We find in particular that the anomalous excitation branch may switch between the spin and charge symmetry sectors without a change in the ground state. Instead it changes its nature by becoming inaccessible to the ground state. 
\end{minipage}
\end{center}

\bigskip\bigskip


%
%
%
%
%

\section{Introduction}

The effort to characterise the behaviour of interacting electrons is the central challenge of condensed matter physics. A key paradigm is Landau's theory of the Fermi liquid, which accounts for the electronic behaviour of a wide variety of systems. It has become clear however that a much richer class of behaviours is possible, with prominent examples being unconventional metals such as pseudogap \cite{Keimer_rev} and heavy fermion \cite{HFrev} systems, and related regimes exhibiting `quanutm critical' fluctuations \cite{Keimer_rev,HFrev}.

An ultimate goal is a complete classification of the possible behaviours of interacting electrons, and to get there a useful strategy is to explore limiting cases. In this work we focus on the special case of correlated electrons in 1d, for which there exists a family of interacting models solvable by Bethe ansatz. While the phenomenology of 1d  is special (in that it differs from generic behaviour in higher dimensions), it can nevertheless be hoped that some insight can be gained which may help clarify the bigger picture.

The nature of low-energy physics in 1d is well understood through the Luttinger liquid framework \cite{Haldane81,GiamarchiLL}, where the long-wavelength excitations are collective modes exhibiting spin-charge separation. For models which are solvable by Bethe ansatz we moreover have access to complete energy spectrum, generated by the dispersion of stable elementary excitations, which scatter non-trivially yet elastically. At half-filling and zero magnetisation these elementary excitations are completely spin-charge separated as spinons and holons. In this work we categorise these excitations and in doing so identify a structure beyond  the reach of the Luttinger liquid framework.

In particular we consider the Hubbard--Shastry family of integrable models of correlated electrons in 1d \cite{HS1}.
Focusing on the half-filled and unmagnetised ground state we find that the elementary excitations admit a four-way classification, depending on whether a gapless branch or an anomalous branch belong to either the spin or the charge sectors. In addition we identify paths in parameter space where the transitions between the distinct excitation patterns can be traced. We pay special attention to the transition of the anomalous branch between the spin and charge symmetry sectors without any change in the corresponding ground state.

The paper is structured as follows. First in Sec.~\ref{sec:gen} we consider a very general lattice model of interacting electrons, discussing symmetries and self-dualities. In Sec.~\ref{sec:sol} we outline the solvable limits giving the Hubbard--Shastry models. The core of the paper is Sec.~\ref{sec:ex} where we discuss the elementary excitations of these models, identifying a four-way classification, and exploring transitions between classes. We conclude in Sec.~\ref{sec:conc}. 
There is one appendix which collects  explicit exact expressions for the elementary excitations employed in Sec.~\ref{sec:ex}.


\section{General lattice model of interacting electrons}\la{sec:gen}

We start with a general lattice model of interacting electrons
\be\la{ham}
\begin{aligned}
\OH=\sum_{\braket{i,j}}\big( \OT_{ij} + \OV_{ij}^s + \OV_{ij}^\eta \big) 
	+ \sum_j \big( U \Oth_j - 2 \mu \Oet^z_j  - h \Os^z_j \big),
\end{aligned}
\ee
with $\braket{\cdot,\cdot}$ denoting summation over nearest neighbour sites. 
Here the kinetic term is of a general correlated form 
\be\la{eq:CH}
\OT_{ij} =t(1-\lam) \OT^\qe_{ij}+t(1+\lam)\OT^\qd_{ij}+t_\pm (\OT^+_{ij}+\OT^-_{ij}),
\ee
where the three parameters $t$, $\lam$, $t_\pm$ decouple the terms 
\be
\begin{split}
\OT^\qe_{ij} &=- \sum_{\s=\ssd,\ssu} \big(\Ocd_{i\sigma} \Oc_{j\sigma} + \Ocd_{j\sigma} \Oc_{i\sigma}\big)\bOn_{i\bs}\bOn_{j\bs},\\
\OT^\qd_{ij} &=- \sum_{\s=\ssd,\ssu} \big(\Ocd_{i\sigma} \Oc_{j\sigma} + \Ocd_{j\sigma} \Oc_{i\sigma}\big)\On_{i\bs}\On_{j\bs},\\
\OT^+_{ij}&=- \sum_{\s=\ssd,\ssu} \big(\Ocd_{i\sigma} \Oc_{j\sigma}\On_{i\bs}\bOn_{j\bs} + \Ocd_{j\sigma} \Oc_{i\sigma}\bOn_{i\bs}\On_{j\bs}\big),\\
\OT^-_{ij}&=- \sum_{\s=\ssd,\ssu} \big(\Ocd_{i\sigma} \Oc_{j\sigma}\bOn_{i\bs}\On_{j\bs} + \Ocd_{j\sigma} \Oc_{i\sigma}\On_{i\bs}\bOn_{j\bs}\big),
\end{split}
\ee
with $\bs=-\s$ and $\bOn_{\s}=1-\On_{\s}$. The spin and charge interactions are given respectively by
\be
\begin{aligned}
\OV_{ij}^s &= J^{zz}_s \Os^z_i\Os^z_j+\tfrac{1}{2} J^{xy}_s \big(\Os^+_i\Os^-_j + \Os^-_i \Os^+_j\big),\\
\OV_{ij}^\eta &= J^{zz}_\eta \Oet^z_i \Oet^z_j+\tfrac{1}{2} J^{xy}_\eta \big(\Oet^+_i\Oet^-_j + \Oet^-_i \Oet^+_j\big),
\end{aligned}
\ee
with 
\be\la{su2s}
\begin{aligned}
  \Os_j^+&=\Ocd_{j\ssu} \Oc_{j\ssd},\qquad \Os_j^-=\Ocd_{j\ssd} \Oc_{j\ssu},\qquad\Os_j^z=\tfrac{1}{2}\big(\On_{j\ssu}-\On_{j\ssd}\big),\\
\qquad \Oet_j^+& =\Ocd_{j\ssd} \Ocd_{j\ssu},\qquad \Oet_j^- =  \Oc_{j\ssu} \Oc_{j\ssd},\qquad\Oet_j^z=\tfrac{1}{2}\big(\On_{j\ssu}+\On_{j\ssd}-1\big).
\end{aligned}
\ee
The onsite term incorporates the Hubbard interaction, with $U$ coupled to
\be
\Oth_j = \big(\On_{j\ssu}-\tfrac{1}{2}\big)\big(\On_{j\ssd}-\tfrac{1}{2}\big),
\ee
as well as a chemical potential $\mu$ and magnetic field $h$.
For the purposes of this section we can generally  consider the model on a bipartite lattice (with $(-1)^j=\pm 1$), with fixed coordination number $z$, in any dimension. In the following sections we then restrict to 1d. 

The model conserves both charge and spin, i.e.~it commutes with the  $U(1)$ symmetry generators  $\Os^z=\sum_j \Os_j^z$ and $\Oet^z=\sum_j \Oet_j^z$. Moreover, for $t_\pm=0$ the model possesses an additional $U(1)$ symmetry generated by $\Oth=\sum_j \Oth_j$, corresponding to the conservation of the total number of doubly occupied sites. For $J^{xy}_s=J^{zz}_s$ the model possesses a dynamical\footnote{We call a symmetry dynamical if its generators, $\OA_\a$ say, obey $[\OH,\OA_\a]=\lambda_\a \OA_\a$ for some scalars $\lambda_\a$. While the $\OA_\a$ are not conserved under Heisenberg's equation of motion $\Der{\OA_\a}{t} = \pDer{\OA_\a}{t} +  \ii [\OH,\OA_\a]$, the dynamical set of generators $\tilde \OA_\a(t) =  e^{-\ii \lambda_\a t} \OA_\a$ are, and the $\tilde \OA_\a(t)$ obey the same algebra as the $\OA_\a$.} spin $SU(2)$ symmetry, obeying
\be
[\OH,\Os^z] = 0,\qquad [\OH,\Os^\pm] = \mp  h \Os^\pm,
\ee
 with $\Os^\pm=\sum_j \Os_j^\pm$. For $J^{xy}_\eta=J^{zz}_\eta$ and $\lam=0$ the model possesses a dynamical charge $SU(2)$ symmetry, obeying
\be
[\OH,\Oet^z] = 0,\qquad [\OH,\Oet^\pm] = \mp   \mu \Oet^\pm,
\ee
 with $\Oet^\pm=\sum_j (-1)^j \Oet_j^\pm$.

In addition, the model possesses a dynamical $SU(2|2)$ symmetry along two distinguished lines in parameter space. 
Firstly, for $t_\pm=\tfrac{1-\k^2}{1+\k^2}t$,  $\lam=0$, $J^{zz}_s = J^{xy}_s =-J^{zz}_\eta = J^{xy}_\eta=\tfrac{4 \k}{1+\k^2} t$,  $U=\tfrac{ 2  \k}{1+\k^2} z t$, the Hamiltonian obeys
\be\la{eq:dym_sym_B}
[\OH, \Oqd_{j\s\nu}] =  \big( zt- \nu \mu -\s h\big) \Oqd_{j\s\nu}, \qquad
[\OH, \Oq_{j\s\nu}] = - \big(zt- \nu \mu -\s h\big) \Oq_{j\s\nu},
\ee
with $\s\in\{\bsd,\bsu\}$ and $\nu\in\{\bqe,\bqd\}$, where
\be\la{eq:qB}
\Oq_{\s\qe} = \sum_j \tfrac{1+\k}{2}\Ocd_{j\bs} - \k \On_{j\s}\Ocd_{j\bs} ,\qquad 
\Oq_{\s\qd} = \sum_j  (-1)^j \s \big(\tfrac{1-\k}{2}\Oc_{j\s} + \k \On_{j\bs}\Oc_{j\s}\big),
\ee
which  obey the algebraic relations
\begin{equation}\label{eq:q_q}
\begin{split}
&  \{ \Oq_{\s\nu}, \Oqd_{\s\nu}\}= \tfrac{1+\k^2}{4}\vol-   \k\, (\s \Os^z - \nu\Oet^z),\\
& \{\Oq_{\ssd\nu}, \Oqd_{\ssu\nu}\}=\k\,{ \Os}^+, ~~~~~~~ 
	\{ \Oq_{\s\qe},\Oqd_{\s\qd}\}= -\k\,{ \Oet}^+,\\
& \{ \Oq_{\ssu\nu}, \Oqd_{\ssd\nu}\}= \k\,{ \Os}^-, ~~~~~~~
	\{\Oq_{\s\qd}, \Oqd_{\s\qe}\}= -\k\,{\Oet}^-,
\end{split}
\end{equation}
where $\s$ takes values $-1,1$ for $\s=\bsd,\bsu$,  and $\nu$ takes values $-1,1$ for $\nu=\bqe,\bqd$, and  $\vol$ is the total number of sites. The second distinguished line is given by $t_\pm=\lam=0$,  $J^{zz}_s = J^{xy}_s =-J^{zz}_\eta = J^{xy}_\eta=2t$, and here the Hamiltonian obeys
\be\la{eq:dym_sym_EKS}
[\OH, \Oqd_{j\s\nu}] =  \big(\tfrac{zt+U}{2}- \nu \mu -\s h\big) \Oqd_{j\s\nu}, \quad
[\OH, \Oq_{j\s\nu}] = - \big(\tfrac{zt+U}{2}- \nu \mu -\s h\big) \Oq_{j\s\nu},
\ee
with the $\Oq$ here given by Eq.~\eqref{eq:qB} evaluated at $\k=1$. In both cases we may regard $\Oq_{\s\qe}$ and $\Oq_{\s\qd}$ as governing exact fermionic excitations of the system, and we emphasise that this is so for a bipartite lattice in any dimension. (On a hypercubic lattice $\Oq_{\s\qe}$ and $\Oq_{\s\qd}$ correspond to exact fermionic excitations of momentum $\vec 0$ and $\vec\pi$ respectively.)

For $\lam=0$ the model is self-dual under the Shiba transformation \cite{shiba}
\be\la{self-duality}
\Ocd_{j\ssd} \leftrightarrow \Ocd_{j\ssd},\quad\Oc_{j\ssd} \leftrightarrow \Oc_{j\ssd},\quad \Ocd_{j\ssu} \leftrightarrow (-1)^j \Oc_{j\ssu},
\ee
which interchanges spin and charge, mapping the model parameters as folllows
\be\la{eq:Shiba}
\{t,t_\pm,J^{zz}_s, J^{xy}_s ,J^{zz}_\eta , J^{xy}_\eta,U,\mu,h\}\leftrightarrow
\{t,t_\pm,J^{zz}_\eta, -J^{xy}_\eta ,J^{zz}_s , -J^{xy}_s,-U,h,\mu\}.
\ee

\section{Solvable limits: the Hubbard--Shastry models}\la{sec:sol}

The model is exactly solvable in any dimension at the non-interacting point 
\be\la{eq:non_int}
t_\pm=t, \quad\lam=J^{zz}_s= J^{xy}_s =J^{zz}_\eta = J^{xy}_\eta=U=0,
\ee
where the spectrum decomposes completely into uncoupled single-particle modes. In 1d there are a number of additional limits where the model is solvable by Bethe ansatz, for which the spectrum can be characterised by the dispersion of stable elementary excitations scattering elastically off one another. We will focus on a family of such integrable limits falling under the name of the Hubbard--Shastry models \cite{HS1}.

\begin{figure}[tb]
\centering
\includegraphics[height=70mm]{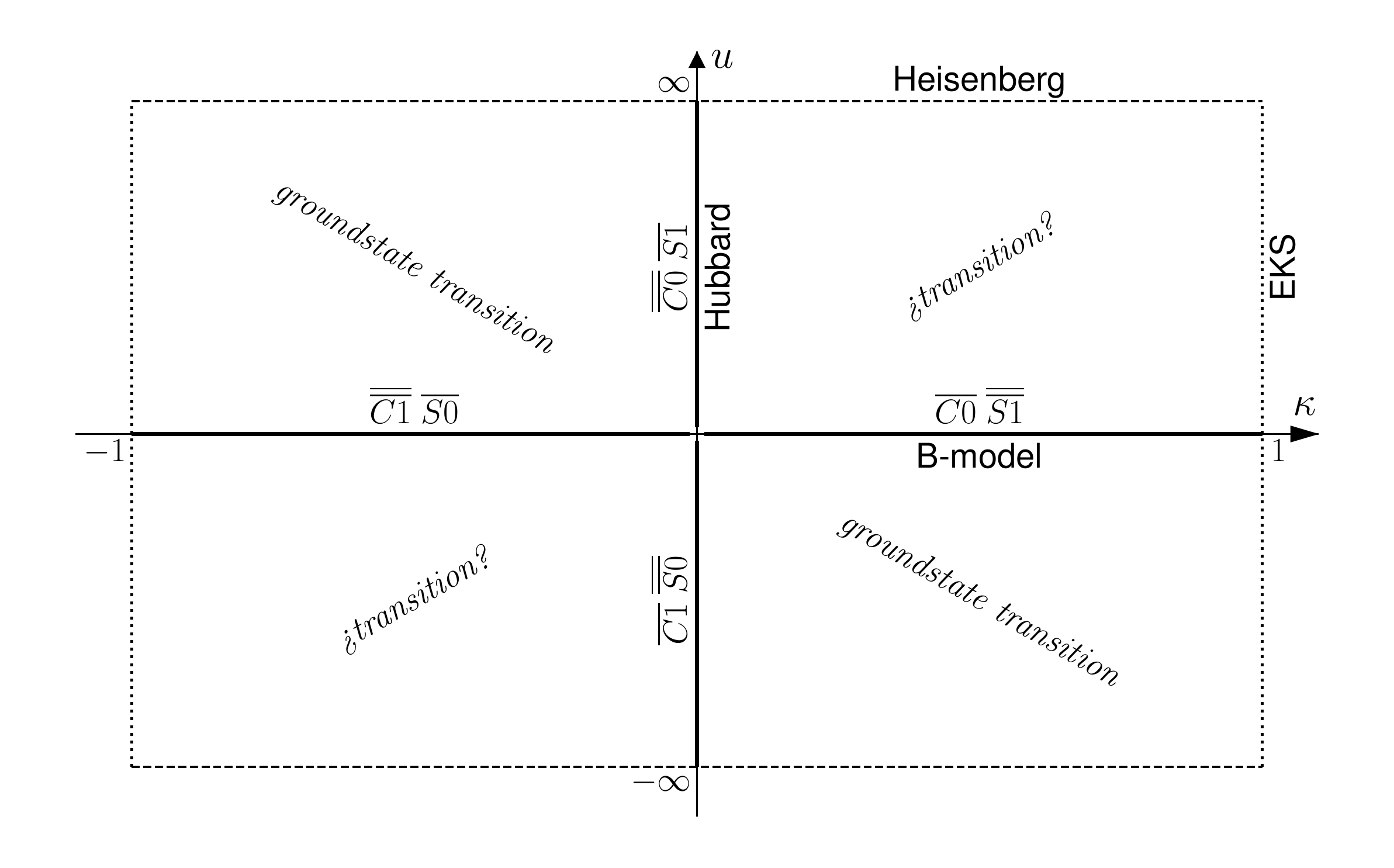}
\caption{\label{diag} 
A schematic depiction of the parameter space determined by Eq.~\ref{eq:param_space}. The structure of the excitations on the integrable axes of the diagram are labelled by $CxSy$, where $x$ and $y$ represent the number of gapless excitation branches in respectively the charge and spin  sectors, and the  overlines denote the number of branches in each sector.  The origin corresponds to the non-interacting point Eq.~\ref{eq:non_int}, from which the physics is non-perturbative.
In the upper-left and lower-right quadrants there is a transition as a gapless branch switches between sectors, which reflects a fundamental change in the ground state. The existence/nature of a transition in the upper-right and lower-left quadrants due to the switching of an anomalous branch between sectors however remains to be understood, as addressed in Sec.~\ref{urq}.}
\end{figure}

To proceed is it useful to introduce a parameterisation of the model parameters as follows
\be\la{eq:param_space}
t_\pm=\tfrac{1-\k^2}{1+\k^2},\quad \lam=0, \quad 
	J^{zz}_s = J^{xy}_s =-J^{zz}_\eta = J^{xy}_\eta = \tfrac{4 \k}{1+\k^2} ,\quad U=u+\tfrac{ 4  \k}{1+\k^2},
\ee
and normalising $t=1$.
This incorporates the following Bethe ansatz solvable models:
\bei
\item The Hubbard chain at $\k=0$, with coupling constant $u$. 
\item The Hubbard--Shastry B-model at $u=0$, with coupling constant $\k$ \cite{HS1}. This possesses the dynamical symmetry given by Eq.~\eqref{eq:dym_sym_B}.
\item The Essler--Korepin--Schoutens model at $\k=1$ \cite{EKS,EKS2}. This possesses the dynamical symmetry given by Eq.~\eqref{eq:dym_sym_EKS}. In addition this possesses the $U(1)$ symmetry generated by $\Oth$, and here the free parameter $u$ can be regarded as a chemical potential for the number of doubly occupied sites. The model at $\k=-1$ is related via the  Shiba transformation Eq.~\eqref{self-duality}.
\item The Heisenberg spin-chain at $u=\infty$. Here the empty and doubly occupied sites become energetically unachievable, leaving only the spin degrees of freedom, and an isotropic antiferromagnetic coupling as the only non-trivial interaction. Conversely at $u=-\infty$ the singly occupied sites become energetically unachievable, giving a `charge-chain' with only the charge-charge interactions determined by $J^{zz}_\eta$, $J^{xy}_\eta$ remaining non-trivial. This is dual to the Heisenberg spin-chain via the Shiba transformation Eq.~\eqref{self-duality}.
\eei
We present a schematic depiction of this parameter space in Fig.~\ref{diag}, which serves to illustrate the four-way classification of excitations discussed in the following section. The Shiba transformation Eq.~\eqref{self-duality} acts as an inversion symmetry about the origin. We collect exact formulae for the dispersion of elementary excitations of these models from Refs.~\cite{Hbook,HS2,EKSTBA}, and present these explicitly in App.~\ref{Adisp}.

\section{Structure of electronic excitations in 1d}\la{sec:ex}

We begin with a general discussion of excitation spectra in 1d, leading to a four-way classification of the elementary excitation of electronic systems above singlet ground states. We then exemplify this with a  discussion of the excitation spectra of the Hubbard--Shastry models.

An important consequence of interactions in the restricted geometry of 1d  is the `fractionalisation' of the underlying degrees of freedom. For example in the case of the antiferromagnetic Heisenberg spin chain the elementary excitations are (spin $\pm\half$) spinons with momentum range $\pi$, as opposed to (spin $\pm1$) magnons with momentum range $2\pi$ \cite{FaddTakh}. Physical excitations are composed of pairs of these elementary excitations.

For the interacting electrons, one has that the electronic, spin and charge degrees of freedom are fractionalised into (spin $\pm\half$, charge $0$) spinons and (spin $0$, charge $\pm1$) holons in an analogous way \cite{exHubb}. Again these elementary excitations having momentum range $\pi$, and physical excitations are composed of a combined even number of the elementary spinon and holon excitations.

There is a subtlety concerning the number of elementary excitation branches. As  an electronic system has four states per site, we may regard the spectrum as emerging from three bands of excitations: two coming from the electronic sector, ``$\{\Oc_{j\s},\Ocd_{j\s}\}=1$'',  and one coming from the spin-charge sector, ``$\{\Os_j^-,\Os_j^+\} + \{\Oet_j^-,\Oet_j^+\}=1$''.
While in the non-interacting limit the Hilbert space essentially decomposes in two, which trivialises the third branch, this is not so for the Bethe ansatz models we consider here. Instead we find that there are three branches of elementary excitations  distributed between the spin and charge sectors, with one of the sectors having an additional branch which we will refer to as the `anomalous' branch\footnote{This is so for the half-filled unmagnetised case on which we focus, where the ground states is a singlet of both the spin and charge $SU(2)$ symmetries. More generally there may exist bound states in the spectrum. These transform under higher-dimensional representations of the symmetries, the generators of which act on more than one site.}.

We further highlight two general facts about the nature of the low-energy excitations in 1d. On the one hand, the generalisation of the Lieb-Schultz-Mattis theorem to electronic systems tells that the ground state is either gapless or degenerate \cite{LSMfermions}. On the other, bosonisation considerations indicate that there cannot be linearly dispersing gapless excitations in both the spin and charge sectors when a model possesses both spin and charge $SU(2)$ symmetries explicitly \cite{AffleckHaldane87}.  From these we can conclude that above a singlet ground state there must be a gapless excitation in one, and only one, sector (spin or charge).

Thus there are four possible ways to characterise the excitation spectra above a singlet ground state for interacting electrons in 1d, depending on whether the gapless or anomalous excitation branch belong to either the spin or charge sector. We adopt a  convenient notation for characterising this structure \cite{CxSy}, e.g.~$\overline{\overline {Cx}}\,\overline{Sy}$, where $x$ and $y$ represent the number of gapless excitations in respectively the charge and spin  sectors, and the  overlines denote the number of branches in each sector (in this example two in the charge sector and one in the spin sector).

The four classes of excitations can be found within the Hubbard and B-models discussed above in Sec.~\ref{sec:sol}, which form the axes of Fig.~\ref{diag}.
\begin{figure}[tb]
\centering
\includegraphics[height=45mm]{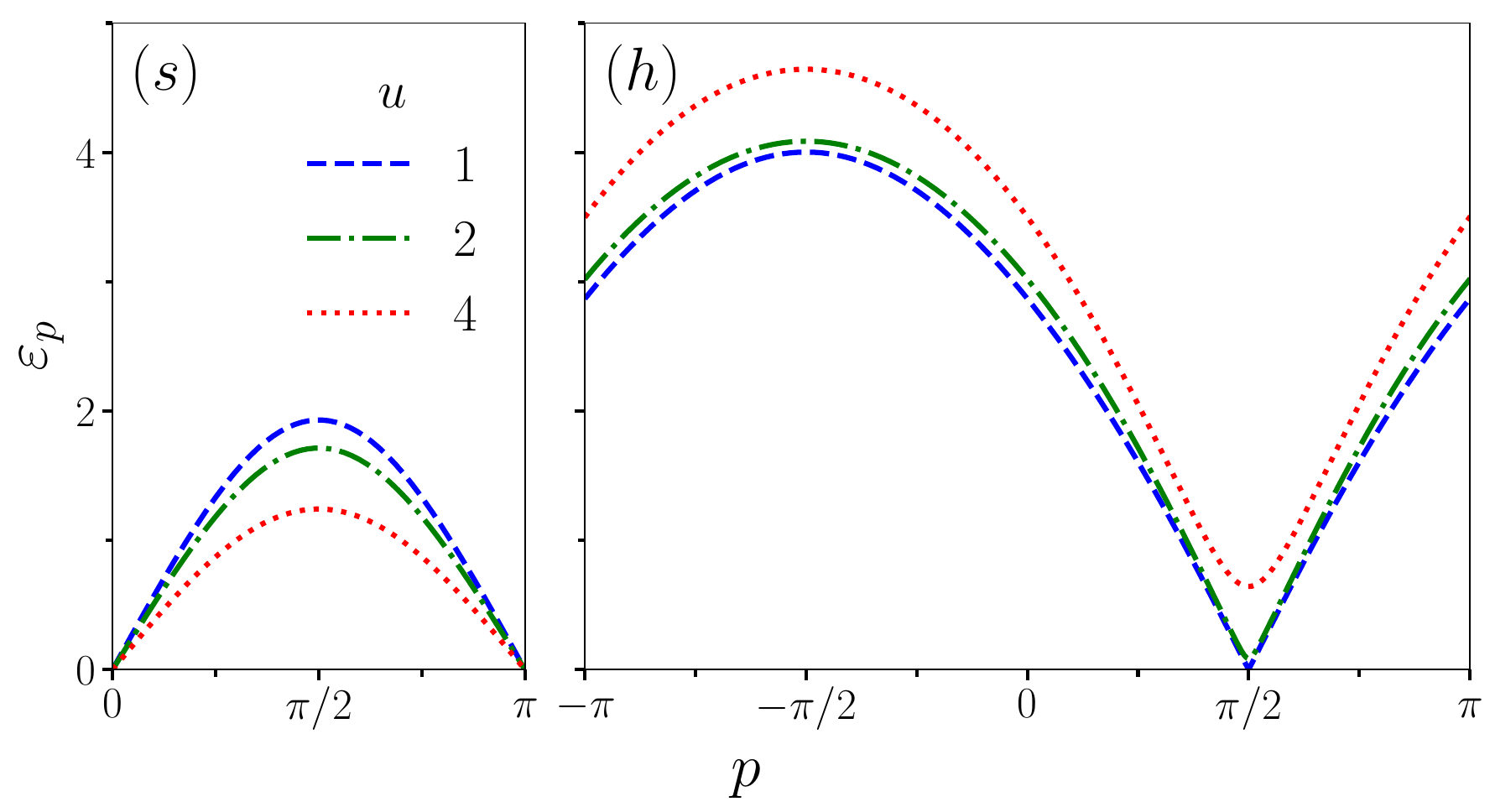}\qquad \includegraphics[height=45mm]{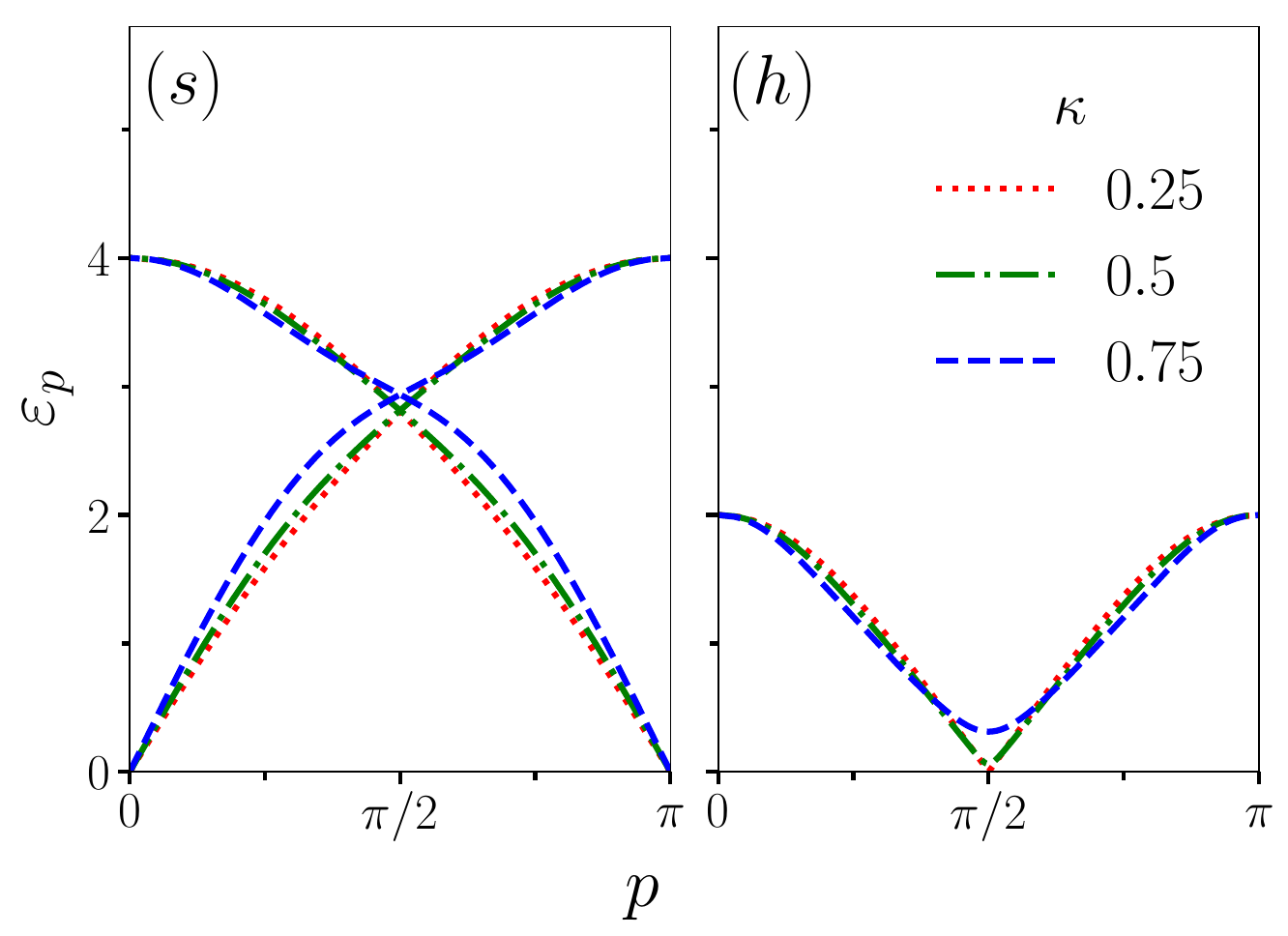}
\caption{\label{HBdisp}
Dispersion curves for the elementary spinon (s) and holon (h) excitations above a half-filled, non-magnetised ground state.
H: the repulsive Hubbard model, for $u=1,2,4$. The spinon momentum extends over a range $\pi$ while the holon momentum extends over a range $2\pi$, indicating two branches in the charge sector. For small $u$ the charge gap is beyond the resolution of the plot. B: the B-model, for $\k=0.25,0.5,0.75$. Here the holon momentum extends over a range $\pi$, while the spinon dispersion curves exhibit an `hourglass' form with two branches over a momentum range $\pi$. Again for small $\k$ the charge gap is beyond the resolution of the plot.
}
\end{figure}
\bei
\item $\overline{\overline {C0}}\,\overline{S1}$: The $u>0$ Hubbard model. The dispersion curves for the elementary excitations are presented on the left side of Fig.~\ref{HBdisp}. The spinons are gapless, whereas there is a Mott gap for the holons for any $u>0$ \cite{LiebWu68}. 
The $\pi$ range of the  spinon momentum indicates a single branch in the spin sector.
In contrast, the $2\pi$ range of the holon momentum indicates  two branches of elementary holon excitations. 
\item $\overline {C0}\,\overline{\overline{S1}}$: The $\k>0$ B-model. Here the dispersion curves for the elementary excitations are presented on the right side of Fig.~\ref{HBdisp}. Like the $u>0$ Hubbard model, the ground state has gapless spinon and gapped holon elementary excitations. The distinction is that here the anomalous branch appears in the spin sector,  giving the distinctive `hourglass' form.
\item $\overline {C1}\,\overline{\overline{S0}}$: The $u<0$ Hubbard model. This is related to the $u>0$ case by the Shiba transformation Eq.~\eqref{self-duality}, which interchanges the spin and charge degrees of freedom.
\item $\overline{\overline {C1}}\,\overline{S0}$: The $\k<0$  B-model. This is related to the $\k>0$ case by the Shiba transformation Eq.~\eqref{self-duality}, which interchanges the spin and charge degrees of freedom.
\eei
In addition we can characterise the elementary excitations around the outer boundary of Fig.~\ref{diag}, and we now do this to analyse the transitions between the classes.

\subsection{Transition of the anomalous branch between sectors}\la{urq}

We now consider the upper-right quadrant of Fig.~\ref{diag} (lower-left quadrant is related by the Shiba transformation) where the low-energy physics is Mott insulating, i.e.~with the charge sector gapped and the spin sector gapless. We wish to focus on the transition of the anomalous excitation branch from the charge sector (for the $u>0$ Hubbard model) to the spin sector (for the $\k>1$ B-model). Passing through the non-interacting point is highly singular, as there the excitations collapse onto uncoupled single-particle modes, and in 1d all physics is non-perturbative from here. Instead we proceed to analyse the elementary excitations on the outer boundary of the quadrant, along which the model remains integrable.

We start from the $u>0$ Hubbard model, for which the anomalous branch belongs to the charge sector. 
Increasing $u$ causes the charge gap to diverge, and in the limit $u\to\infty$ the physics is projected onto the spin degrees of freedom $\ket{\ssd},\ket{\ssu}$. That is, along the upper boundary of the quadrant the model is the antiferromagnetic Heisenberg spin chain with $|J|=\tfrac{4 \k}{1+\k^2}$, and only the spinon excitation branch remains in the accessible spectrum.  

\begin{figure}[tb]
\centering
\includegraphics[height=50mm,width=90mm]{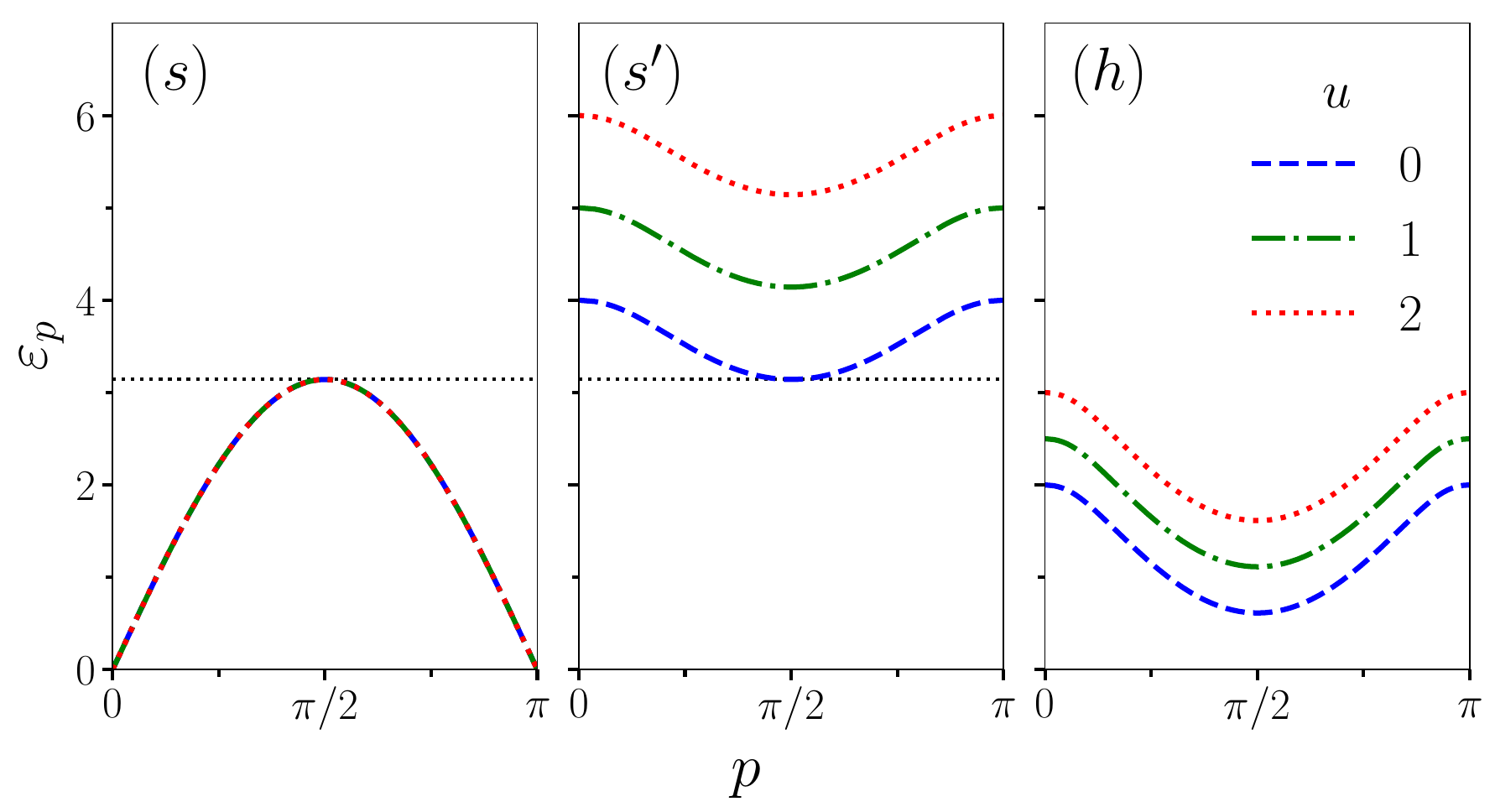}
\caption{\label{EKSdisp}
Dispersion curves for the elementary excitations of the EKS model, for $u=0,1,2$, above the singlet half-filled unmagnetised ground state. 
In addition to the low-energy spinon branch (s), there is a holon branch (h) and an anomalous spinon branch (s'), both of which are charged under $\Oth$.  Varying $u$ shifts the energy of the (h) and (s') branches, while leaving the (s) branch invariant. A horizontal line is drawn in the first two panels to guide the eye to the touching of the two spin branches when $u=0$, which coincides with the $\k\to1$ limit of the B-model.
}
\end{figure}

We can analyse the reemergence of the projected degrees of freedom for finite $u$ at $\k=1$, where we have the EKS model. 
The elementary excitations here are presented in Fig.~\ref{EKSdisp}, revealing that the returning branches are split between the spin and charge symmetry sectors, with one in each. 
These two branches are furthermore charged under $\Oth$, which as highlighted above generates an additional $U(1)$ symmetry for the EKS model. As a consequence, varying $u$ here shifts these branches through the spectrum without affecting the corresponding eigenstates, and in particular the ground state remains precisely that of the antiferromagnetic Heisenberg chain for $u>-4+4\log 2\approx-1.23$ \cite{EKS2}.  
Let us remark that the $\pi$ momentum shift of between the anomalous  spinon branch here relative to the anomalous holon branch of the ($u>0$) Hubbard model can be attributed to the spin operators $\Os^\pm$ having momentum $0$, while the   charge operators $\Oet^\pm$ have momentum $\pi$.
At $u=0$ the bottom of the anomalous spinon branch touches the top of the low-energy spin branch, and upon moving to the B-model line the $U(1)$ symmetry generated by  $\Oth$ gets broken and the two spinon branches hybridise in the `hourglass' form of Fig.~\ref{HBdisp}.

To summarise, the ground state of the model has remained exactly identical along the outer boundary of the quadrant as the anomalous branch switched between the spin and charge sectors. This in stark contrast to a widely held perspective that a system's excitations are intimately linked to its ground state, and cannot change unless an excitation gap closes. Instead what we have seen is that the gap for the anomalous branch diverged, making it inaccessible to the ground state, and thereby allowing it to change its nature.

More generally, within the interior of the quadrant the integrability of the model gets broken, inducing decay of elementary excitations, and the  spectrum can no longer be  characterised exactly. We may nevertheless conjecture that the anomalous branch cannot be continuously traced between the Hubbard and B-models along a generic path passing through the interior of the quadrant. It is natural to assume that it must at some point become inaccessible to the ground state along such a  generic path, as it does at both at the outer boundary and at the non-interacting point. We leave as an interesting direction for future work to explore whether the required breakdown of the anomalous excitation branch leaves a signature which may be used to characterise it further.

\subsection{Transition of the gapless branch between sectors}

We now turn to the upper-left quadrant of Fig.~\ref{diag} (lower-right quadrant is related by the Shiba transformation).  
Here a gapless branch switches between sectors, and so a change in the ground state is guaranteed.

We proceed by tracing the boundary of the upper-left quadrant, this time proceeding in an anti-clockwise direction. 
Upon taking the $u\to\infty$ limit of the Hubbard model, one obtains the $|J|=\tfrac{4 |\k|}{1+\k^2}$ ferromagnetic Heisenberg spin chain for $\k<0$. Here the excitations are quadratically dispersing spin $\pm1$ magnons and their bound states. Decreasing $u$ along the $\k=-1$ line brings back the two additional branches,  which are again charged under $\Oth$. Both however have non-dispersive flat bands. One is an electronic excitation, which is allowed here as the degenerate ground state breaks spin-charge separation, and the other carries spin 1.  When $u$ becomes $4$, these flat bands touch the ground state, and induce a sudden change in the nature of excitations\footnote{This highly singular point also corresponds to a limit of the Hubbard-Shastry A-model, which for finite interaction parameter describes itinerant ferromagnetism in 1d \cite{HS1,HS2}.}. As $u$ is further lowered the ground state becomes ferrimagnetic\footnote{As we restrict attention to the zero magnetisation sector, the ground state is at the centre of a degenerate multiplet, the size of which decreases along this interval.}, down as far as $u=4-4\log 2\approx 1.23$, and upon reaching this value the excitations become again  spinons and holons, with the anomalous branch reappearing in the charge sector. Decreasing $u$ further the spinons acquire a gap and the holons linearly disperse, and this behaviour continues along the $\k<0$ line of the B-model. 

This  analysis demonstrates the manner in which the ground state becomes degenerate in this quadrant, as must happen to be compatible with the two general facts  on the nature of the low-energy excitations outlined above.


\section{Conclusion}\la{sec:conc}

Focusing on the solvable limits of a general nearest-neighbour Hamiltonian, we have revealed a hidden structure in the spectra of 1d electronic lattice models. Through this we have obtained a four-way classification of the excitations above a  half-filled unmagnetised singlet ground state, as summarised in Fig.~\ref{diag}. We have provided examples for each class through the Hubbard--Shastry models, and have discussed transitions between the classes. 

We have paid particular attention to the switching of an anomalous excitation branch between the spin and charge sectors, which occurs without the closure of a gap. Specifically we have identified a path along which the switching of the branch can be traced exactly, and along which the ground state remains exactly identical. We have described this as an unconventional kind of `transition', where the anomalous excitation becomes inaccessible to the ground state. More generally, this may be regarded as a complete breakdown of the quasi-particles corresponding to the anomalous branch of excitations.

\section*{Acknowledgements}
We thank  S.~Frolov and E.~Ilievski  for valuable discussions. Support from the Foundation for Fundamental Research on Matter (FOM),  the Netherlands Organization for Scientific Research (NWO), and the European Research Council under ERC Advanced grant 743032 DYNAMINT is gratefully acknowledged.

\begin{appendix}

\section{Dispersion curves}\la{Adisp}

In this appendix we present the dispersion relations for the excitations of the model along its exactly solvable lines, which can be found, up to conventions, in \cite{Hbook,HS2,EKSTBA}. These are sufficient to reproduce the plots in Figs.~\ref{HBdisp} and \ref{EKSdisp}.
 We will restrict attention to excitations above a half-filled, zero-magnetised ground state, but we shall present dependence on $h$ and $\mu$, so that the dressed spin and charge of the excitations can be read off from $-\pDer{E}{h}$ and $-\pDer{E}{\mu}$.

Before proceeding let us first introduce some useful conventions. We denote convolutions between kernels $K(v,t)$ and functions $f(v)$ as
\be
\begin{aligned}
\big(K\st f\big)(v) = \int_{-\infty}^\infty dt\ K(v,t) f(t),\quad
\big(K\hst f\big)(v) = \int_{-1}^1 dt\ K(v,t) f(t),\quad
K\cst f=K\st f-K\hst f.
\end{aligned}
\ee
Some useful functions are
\be
\begin{aligned}
&s(v) = \frac{1}{4c \cosh  \frac{\pi v}{2c}},\quad
K_M(v) = \frac{M}{\pi c(M^2+v^2/c^2)},\quad
\Theta_M(v)=2\arctan\big(\frac{v}{c M}\big),\\
&\Upsilon(v)= i\log\Big[\frac{\Gamma(\frac{1}{2}+ \frac{iv}{4 c}) \Gamma(1- \frac{iv}{4 c}) }{  
 	\Gamma(\frac{1}{2}- \frac{v}{4 c}) \Gamma(1 +  \frac{iv}{4 c}) }\Big] \,, \quad
 \Psi(v)= \frac{\pi}{ 2} - 2 \arctan \Big[ \exp \big( \frac{\pi v}{ 2 c} \big)\Big]\,.
\end{aligned}
\ee  
A function appearing on the left side of a convolution is regarded  as a  two-variable kernel as follows $K(v,v') \equiv K(v-v')$.

\subsection{Hubbard model with $u>0$}

The energy and momenta of the spinon and holon excitations of the Hubbard model are given parametrically by 
\be
\begin{aligned}
&E_s = s \hst (e_+-e_-) -h/2, \quad
&&P_s = - \frac{1}{2\pi}\Psi\hst\big(\Der{p_+}{v}-\Der{p_-}{v}\big)+\pi/2 , \\
&E_h^\pm = - e_\pm +s \st e_\B -\mu , \quad
&&P_h^\pm =   p_\pm - \frac{1}{2\pi}\Upsilon\hst\big(\Der{p_+}{v}-\Der{p_-}{v}\big)-\pi/2 , 
\end{aligned}
\ee
where $E_h^\pm$ and $P_h^\pm$ take values on the interval $(-1,1)$, and $p_\pm(v) = \frac{1}{i}\log(iv\mp\sqrt{1-v^2})$, $e_\pm(v)=-2\cos p_\pm(v)-2c$, $e_\B(v) =e_+(v+ic)+e_+(v-ic)$, and $c=u/4$.

\subsection{B-model with $\k>0$}

The energy and momenta of the spinon and holon excitations of the B-model are given parametrically by (correcting a shift by $\pi/2$ typo in Eq. (3.14) of \cite{HS2})
\be\la{Bex}
\begin{aligned}
&E_s^\pm = e_\pm -s \st e_\B - h/2, \quad
&&P_s^\pm =   p_\pm - \frac{1}{2\pi}\Upsilon\cst\big(\Der{p_+}{v}-\Der{p_-}{v}\big)-\pi/2 , \\
&E_h =-s \cst (e_+-e_-) -\mu, \quad
&&P_h =  -\frac{1}{2\pi}\Psi\cst\big(\Der{p_+}{v}-\Der{p_-}{v}\big)+\pi/2, 
\end{aligned}
\ee
where $E_s^\pm$ and $P_s^\pm$ take values on the interval $(-\infty,-1)\cup(1,\infty)$, 
and 
\be
p_\pm(v) = \frac{1}{i}\log\Big(\frac{-\sqrt{1+c^2}\pm iv\sqrt{1-1/v^2}}{c+iv}\Big),
\ee
$p_\B(v) =p_+(v+ic)+p_+(v-ic)$, $e_\pm(v)=-2\cos p_\pm(v)$, $e_\B(v) =e_+(v+ic)+e_+(v-ic)$, and $c=\tfrac{2 \k}{1-\k^2}$.

\subsection{EKS model}

We will break up the description of excitations of the EKS model to focus on the physically most interesting case $u>0$ for the Hamiltonian of Eq.~\eqref{ham}, with the negative $u$ case following from the Shiba transformation Eq.~\eqref{self-duality}.

First focusing on the region $u>-4+4\log2\approx -1.23$ of the model on the $\k=1$ line, 
the energy and momenta of the spinons (s) and (s'),  and  holon (h) excitations are given parametrically by
\be
\begin{aligned}
&E_s = 4\pi s - h/2 , \quad
&&P_s = \Psi  +\frac{\pi}{2} , \\
&E_{s'} = 4-4\pi K_2 \st s - h/2 + u , \quad
&&P_{s'} =  \Psi +\Theta_1 +\pi/2, \\
&E_h = 2- 4\pi K_1 \st s  - \mu+u/2, \quad
&&P_h = -\Upsilon +\pi/2, 
\end{aligned}
\ee
with $c=1$. In the limit $u\to\infty$ the (s') and (h) excitations disappear from the spectrum, and just the spinon excitation  (s) of the $J=2$ antiferromagnetic XXX chain remains.

Next we turn to the $u>4$ region of the $\k=-1$ line. Here the dispersive excitations are the magnons of the $J=2$ ferromagnetic XXX chain, and their bound states ($M=1,2,3,\ldots$), and their dispersion is given parametrically by
\be
\begin{aligned}
E_M &= 4\pi K_M - Mh , \quad
P_M = \pi - \Theta_M,
\end{aligned}
\ee
with $c=1$. In addition there are electronic (e) and magnonic (1') non-dispersive flat bands whose energy depends on $u$ as
\be
E_e=u/2-2-\mu-h/2,~~~ E_{1'} = u-4-h.
\ee

Below $u=4-4\log 2\approx 1.23$ on the $\k=-1$ line, the excitations can be obtained from Eqs.~\eqref{Bex} through the Shiba transformation \eqref{self-duality}.

Along the region $4-4\log 2<u<4$ on the $\k=-1$ line, the nature of the excitations varies. Their energies correspond to the solution of the coupled system of linear integral equations
\be
\begin{aligned}
&E_1 = 4\pi s - \mu + s\st_{\bar Q} E_{2} , \qquad E_{1'} = 4-u - \mu -4\pi K_2\star s + s\st_{\bar Q} E_{2} , \\
&E_{M+1} = -4+u-Mh + 4\pi K_{M+1} -K_{M+1}\st_{Q} E_{2}-K_{M-1}\st_{Q} E_{2},
\end{aligned}
\ee
with $c=1$, and  $M$ takes values on the positive integers. The subscript on the convolution indicates the interval over which the integral is taken, with $Q=(-\infty,-q)\cup(q,\infty)$, and $\bar Q$ is its complement in $\mathbb{R}$. The parameter $q$ is determined self-consistently by $E_{2}(q)=0$. The spectra should be evaluated at $\mu=h=0$, but keeping these explicit allows one to track the spin and charge of the excitations in this interval. At $u=4-4\log 2$, the holon bands are (1) and (1'), the spinon band is (2), and the higher $(M)$ all have zero energy, and represent the degeneracy arising from the ground state becoming symmetric under the  spin $SU(2)$.

\end{appendix}

\setstretch{1.1}
\setlength{\parskip}{0em}

\bibliographystyle{JHEP}
\bibliography{HSbib}

\end{document}